\pdfoutput=1 
\documentclass{JINST}
\let\ifpdf\relax

\title{ArDM: first results from underground commissioning}

\author{A. Badertscher$^a$, F. Bay$^a$, N. Bourgeois$^b$, C. Cantini$^a$, A. Curioni$^a$, M. Daniel$^c$, U. Degunda$^a$, S. Di Luise$^a$, L. Epprecht$^a$, A. Gendotti$^a$, S. Horikawa$^a$, L. Knecht$^a$, D. Lussi$^a$, G. Maire$^b$, B. Montes$^c$, S. Murphy$^a$, G. Natterer$^a$, K. Nikolics$^a$, K. Nguyen$^a$, L. Periale$^a$, S. Ravat$^b$, F. Resnati$^a$\thanks{Corresponding author.}, L. Romero$^c$, A. Rubbia$^a$, R. Santorelli$^c$, F. Sergiampietri$^a$, D. Sgalaberna$^a$, T. Viant$^a$, S. Wu$^a$\\
\llap{$^a$}ETH Zurich, Institute For Particle Physics\\Schafmattstrasse 20, 8093 Zurich, Switzerland\\
\llap{$^b$}CERN, PH/DT group\\1211 Gen\`eve 23, Switzerland\\
\llap{$^c$}CIEMAT, Divisi\'on de F\'isica de Part\'iculas\\Avenida Complutense, 40, E-28040, Madrid, Spain\\
E-mail: \email{filippo.resnati@cern.com}}

\abstract{The Argon Dark Matter experiment is a ton-scale double phase argon Time Projection Chamber designed for direct Dark Matter searches.
It combines the detection of scintillation light together with the ionisation charge in order to discriminate the background (electron recoils) from the WIMP signals (nuclear recoils).
After a successful operation on surface at CERN, the detector was recently installed in the underground Laboratorio Subterr\'aneo de Canfranc, and the commissioning phase is ongoing.
We describe the status of the installation and present first results from data collected underground with the detector filled with gas argon at room temperature.}

\keywords{Dark Matter; WIMP; Liquid argon; Double phase; TPC}

\begin{document}

\section{Introduction}
Astronomical observations suggest the presence in the Universe of a non-luminous and non-baryonic mass excess, presumably composed of a new type of elementary particle.
The so-called Weakly Interacting Massive Particle (WIMP)~\cite{steigman:1985} is a leading candidate to describe the excess.
WIMPs interact only weakly and gravitationally, and they are expected to scatter elastically on ordinary matter producing nuclear recoils with energies up to few tens of keV~\cite{lewin:1996}.
Argon, in its liquid form, represents a promising target, with a favourable form factor and only sensitive to spin-independent interactions.
Assuming a WIMP mass of 100~GeV/c$^2$ and the WIMP-nucleon cross section of 10$^{-8}$~pb, the expected rate of nuclear recoil above 30~keV in 1~ton of argon is about 0.5~event/day.
Argon is an attractive medium also because it is easily commercially available and cheap, easy to handle and to purify, it is an excellent scintillator, and ionisation electrons can freely drift under the action of an externally applied electric field.
Ionising radiation leads to the formation of excited molecular states in either singlet or triplet states, which decay radiatively with fast and slow characteristic times that are largely different, allowing to easily discriminate between the two components.
The relative abundance of singlet and triplet states depends on the type of ionising radiation~\cite{hitachi:1983}, providing a tool for the background rejection.
In addition, the phenomenon of charge recombination, also dependent on the type of radiation (ionisation density), transforms ionisation electrons into scintillation light~\cite{kubota:1978}, giving yet another powerful discriminant of nuclear recoils (induced by neutrons or WIMPs) against electron recoils (induced by gamma and beta decays).
The ionisation charge arising from a few keV nuclear recoil needs a stage of amplification before being read out.
This is obtained extracting the electrons from the liquid into the vapour argon (hence \emph{double phase} technology)~\cite{dolgoshein:1973} where, by means of an intense enough electric field, drifting electrons can further excite atoms producing secondary scintillation proportional to the amount of primary electrons~\cite{suzuki:1979}.
In vapour argon it is also possible to directly amplify the charge through the Townsend avalanche in high electric field produced in the Large Electron Multiplier (LEM)~\cite{badertscher:2011}, a glass epoxy plate, metal cladded on both sides and perforated in a regular pattern.

The Argon Dark Matter (ArDM) experiment~\cite{rubbia:2006} is using a ton-scale liquid argon target, operated as a double-phase time projection chamber with imaging and calorimetric capabilities.
In the following sections we will describe the detector, the first underground test of ArDM filled with pure argon gas at room temperature, and we will give an estimation of the expected light yield for the liquid argon operation at zero electric field.

\section{The detector}
\begin{figure}[tbp]
\centering
\includegraphics[width=.80\textwidth]{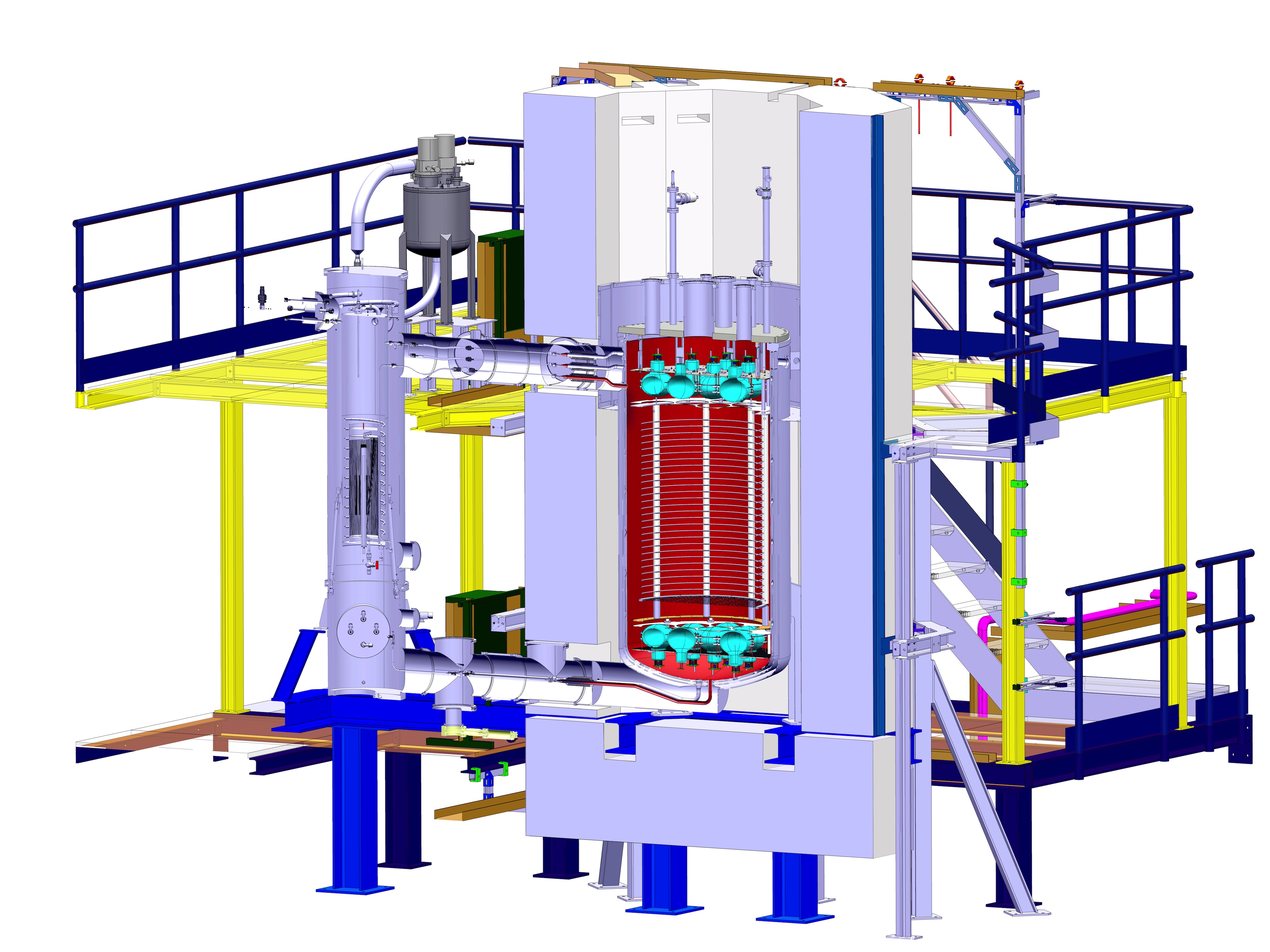}
\caption{CAD design of the ArDM detector and related infrastructures.}
\label{fig:ArDM}
\end{figure}

The ArDM experiment is a liquid argon Time Projection Chamber (TPC) with a vapour layer on top of the liquid allowing a double phase charge readout.
It was constructed and operated on surface at CERN~\cite{boccone:2009} and transferred underground in the Laboratorio Subterr\'aneo de Canfranc (LSC) starting from early 2012.
The actual detector was further improved and updated to make it more suitable for the underground operation.
Figure~\ref{fig:ArDM} shows the CAD design of ArDM and related infrastructures.
The active volume is a cylinder 110~cm long and with a diameter of 80~cm.
The uniformity of the electric field is ensured by a series of stainless steel field shaping rings outside the fiducial volume.
The proper potential to each ring is provided via a resistor chain.
The cathode, at the bottom of the fiducial volume, consists of a transparent stainless steel mesh.
Analogously, two grids made out of parallel 100~$\mu$m diameter stainless steel wires, delimit the fiducial volume at the top.
These grids provide an electric field high enough to efficiently extract the drifting electrons into the vapour phase and produce the secondary scintillation.
The light readout consists of two arrays of twelve 8" Photomultiplier Tubes (Hamamatsu 5912-02MOD-LRI) arranged in honeycomb pattern.
One array is installed below the cathode immersed in liquid argon and the other one in the vapour phase above the extraction grids.
The latter is meant to detect both the primary and secondary scintillation.
In order to efficiently detect the scintillation light a reflector (Tetratex foils supported by Vikuiti Enhanced Specular Reflector) is installed on the field shaping rings, defining the side of the active volume.
Liquid argon scintillation is peaked at 128~nm, making it difficult to be directly detected and reflected.
For this reason a wavelength shifter (1,1,4,4-tetraphenyl-1,3-butadiene -- TPB) is deposited on the PMT windows and on the reflector via vacuum evaporation.
TPB absorbs the light in the VUV region and emits photons with wavelength peaked at 420~nm~\cite{burton:1973}.
Since the TPB deteriorates when exposed to humidity and UV light, the wavelength shifter coatings were renewed before the installation of the detector in LSC.
The thermodynamic conditions in the main vessel are controlled with two cryocoolers that re-condense the boil off of the liquid argon bath that surrounds the main vessel.
In order to ensure the argon purity for the electron drift a recirculation circuit is present: a cryogenic bellows pump pushes the liquid through a copper cartridge, which captures oxygen molecules\footnote{Oxygen, being an electronegative molecule, attaches to drifting electrons resulting in an effective loss of primary charge.}.
A neutron passive shielding, consisting of 50~cm thick polyethylene walls, surrounds the main vessel\footnote{Presently the neutron shielding is only partially installed in order to have easy access to the detector.}.
The slow control processes (measurement of the vacuum, temperatures, pressures; control of the liquid argon recirculation, cooling and level; steering the high voltages) are monitored and managed by a a Programmable Logic Controller (PLC).
For a more detailed report on the status of ArDM one can refer to~\cite{badertscher:2013}.

\section{Underground gas test}
First data taking at LSC was carried out for two weeks in April 2013 operating the detector filled with pure argon gas at room temperature.
The aim of the test was to evaluate the light yield and the background in deep underground environment.
A low activity $^{241}$Am alpha source\footnote{This source provides also 60~keV photons. Their interaction rate in the detector active volume was estimated to be less than 1~Hz. They are neglected in this context.} was installed inside the active volume.
The source could be moved vertically from the cathode up to the extraction grids along the axis of the cylinder.
The 5.5~MeV alpha particles provide a well defined amount of scintillation photons in a localised position, so that the light yield as a function of the drift (Z)~coordinate can be measured.
The data were taken for 5--7 hours/day, limited by the degradation of the purity of the gas argon due to the outgassing of the component of the detector\footnote{A gas purification circuit was not installed at the time of the data acquisition.}.
Therefore, the detector was evacuated during the night and filled with pure gas the next day.
The argon quality was constantly monitored from the characteristic time of the scintillation~\cite{amsler:2008}, and the effect is taken into account in the analysis.
A VME-based electronics was designed for the data taking in gas.
The signals from the 24 PMTs are individually recorded by four CAEN V1720 digitisers (12~bit, 250~MHz).
For triggering, the signals of each PMT array are analog summed, and the coincidence within 100~ns of at least two photoelectrons for both PMT arrays is required.
Dedicated runs for the PMT calibration were taken using LEDs.
The single photoelectron distribution is also obtained from the isolated peaks in the signal tails.
The variation of the gain among the PMTs was within $\pm5$\% for the entire data-taking period.

\section{Results}
\begin{figure}[tbp]
\centering
\includegraphics[width=.80\textwidth]{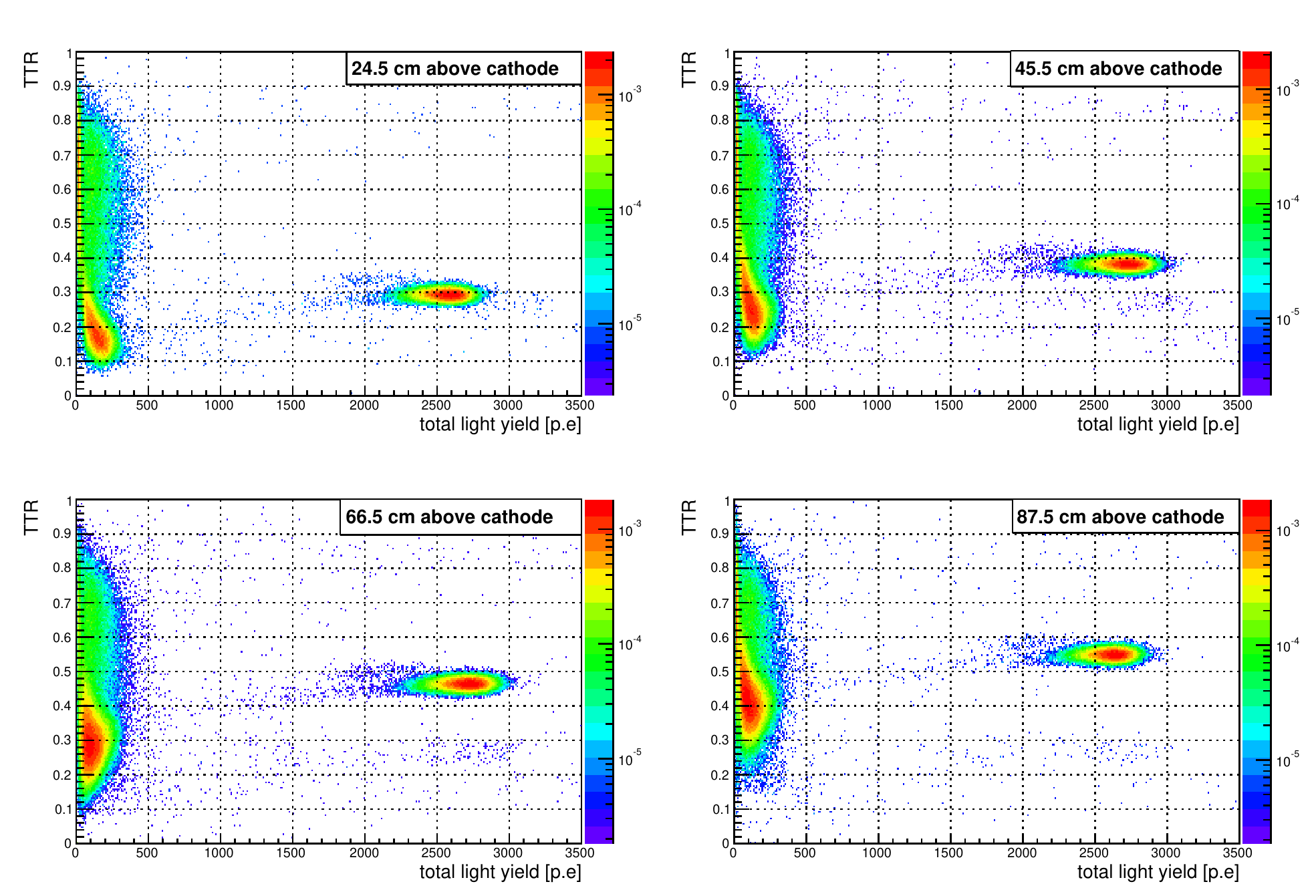}
\caption{Top to total ratio as a function of the total light yield for different positions of the $^{241}$Am source.}
\label{fig:TTR}
\end{figure}

In the following analysis we define the light yield (LY) as the number of photoelectrons contained in both the fast and the slow components of the signal\footnote{The scintillation characteristic time and the amount of light depend on the argon purity~\cite{amsler:2008}. In the analysis we correct for this effect extrapolating the light yield to \emph{infinite} purity. The correction is of the order of 10--30\%.}, i.e.\ the integral over 7~$\mu$s of each event's waveform divided by the integral of a single photoelectron pulse.
We distinguish the light detected by the top PMT array (LY$_{top}$) and the bottom one (LY$_{bot}$), and we define the top to total ratio as TTR~=~LY$_{top}$/(LY$_{top}$ + LY$_{bot}$).
Assuming that the scintillation light is emitted isotropically, the TTR is a measure of the Z~coordinate of the source.
Fig.~\ref{fig:TTR} shows the TTR versus the total LY for different source heights.
Taking the top left plot as an example, one can identify three regions:
\begin{itemize}
\item Events with LY~>~2000~p.e., which are due to the alpha particles depositing their full energy in the gas argon.
\item Events with TTR~<~0.3 and LY~<~500~p.e., which are due to the alpha particles hitting the source holder and, therefore, depositing only a fraction of the energy in the gas argon\footnote{The source holder biasses the light propagation: the isotropic approximation is not valid, and the scintillation preferably shines towards the bottom PMT array. We refer the reader to~\cite{badertscher:2013} for more details.}.
\item Events with TTR~>~0.3 and LY~<~500~p.e., which are due to background events distributed all over the active volume.
\end{itemize}

\begin{figure}[tbp]
\centering
\includegraphics[width=.80\textwidth]{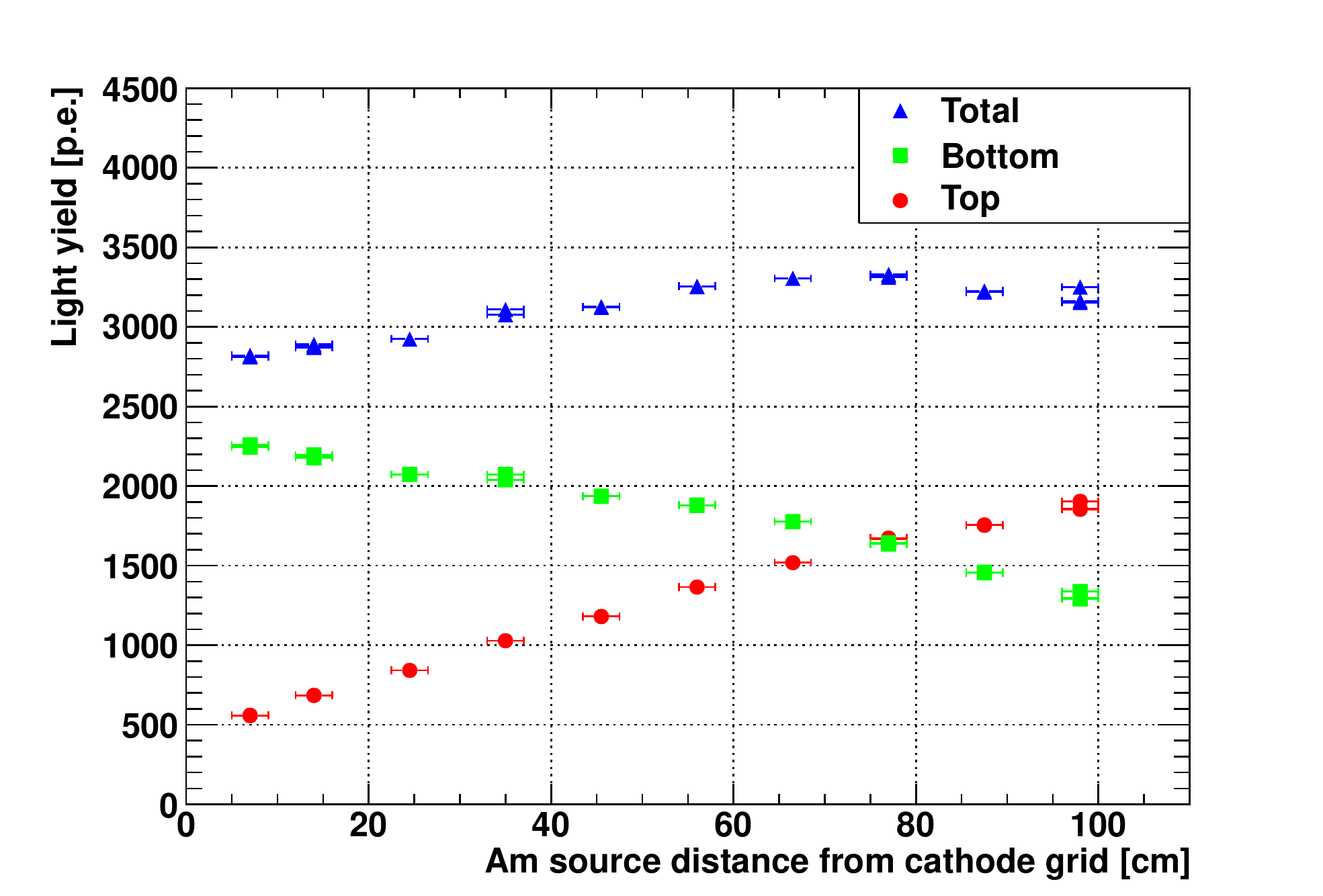}
\caption{Measured light yield extrapolated to \emph{infinite} purity (correction of the order of 10--30\%) for the top PMT array (red), the bottom PMT array (green) and their sum (blue), as a function of the distance between the cathode grid and the $^{241}$Am source.}
\label{fig:pos}
\end{figure}

Moving the source upwards the TTR of the full energy alpha events increases, i.e.\ more light is collected on the top PMT array and less on the bottom one.
The total collected light varies modestly (within $\pm$10\%) with the source position, as shown in Fig.~\ref{fig:pos}.
On average the bottom PMT array detects more light.
This is due to the presence of the source holder shadowing partially the top PMT array.

The light yield of the newly assembled detector improved notably comparing with a similar test performed on surface at CERN in analogous conditions\footnote{The top PMT array was not present at the time.}.
At the time the light yield was at most 850~p.e. and significantly dependent on the position of the source, while in the present test we detect about 3000~p.e..
The alpha particles from the $^{241}$Am source used at CERN had the energy peaked at 4.8~MeV, due to a sealing with a palladium layer, in contrast to the present source that delivers alphas with energy of 5.5~MeV.
Taking this into account, the gain in the light yield is larger by a factor of three.
Using 511~keV photons from a $^{22}$Na source in a previous liquid argon test at CERN with no electric field we measured a light yield of 0.7~p.e./keV$_{ee}$~\cite{lazzaro:2012}.
Assuming that the improvement in light yield will be unchanged in liquid argon\footnote{The source holder, that shadows the top PMT array, will not be present in a liquid argon run, therefore the light yield improvement may be better than a factor of three.}, we expect at least 2~p.e./keV$_{ee}$ for the actual detector in liquid argon with no electric field.

\section{Conclusions}
The ArDM detector and its infrastructure have successfully been installed underground at Laboratorio Subterr\'aneo de Canfranc.
Data collected with ArDM filled with gas argon at room temperature have allowed the assessment of the performance of the newly rebuilt light detection system.
We observe a significant improvement of the light yield and light collection uniformity with respect to the detector configuration tested on surface.
We estimate a light yield of more than 2~p.e./keV$_{ee}$ with zero electric field in liquid argon operation.


\begin{thebibliography}{9}

\bibitem{steigman:1985}
G. Steigman and M. S. Turner,
\emph{Nucl. Phys.} B 253 (1985) 375.

\bibitem{lewin:1996}
J. D. Lewin and P. F. Smith,
\emph{Astropart. Phys.} 6 (1996) 87.

\bibitem{hitachi:1983}
A. Hitachi et al.,
\emph{Phys. Rev.} B27 (1983) 5279.

\bibitem{kubota:1978}
S. Kubota et al.,
\emph{Phys. Rev.} B 17 (1978) 2762.

\bibitem{dolgoshein:1973}
B. Dolgoshein et al.,
\emph{Sov. J. Particles Nucl.} 4 (1973) 70.

\bibitem{suzuki:1979}
M. Suzuki and S. Kubota,
\emph{Nucl. Instr. Meth.} 164 (1979) 197.

\bibitem{badertscher:2011}
A. Badertscher et al.,
\emph{Nucl. Instr. Meth. in Phys. Res.} A 641 (2011) 48.

\bibitem{rubbia:2006}
A. Rubbia,
\emph{J. Phys. Conf. Ser.} 39 (2006) 129.

\bibitem{boccone:2009}
V. Boccone et al.,
\emph{JINST} 4 (2009) P06001.

\bibitem{burton:1973}
W. M. Burton and B. A. Powell,
\emph{App. Opt.}, 12 (1973) 87.

\bibitem{badertscher:2013}
A. Badertscher et al.,
arXiv:1307.0117v1.

\bibitem{amsler:2008}
C. Amsler et al,
\emph{JINST} 3 (2008) P02001.

\bibitem{lazzaro:2012}
C. Lazzaro,
Diss., ETH Zurich, Nr. 20551 (2012).

\end{thebibliography}
\end{document}